\documentclass[twocolumn,floatfix,showpacs,amsmath,amssymb]{revtex4}
\usepackage{graphicx}

\newcommand\beq{\begin{equation}}
\newcommand\eeq{\end{equation}}
\newcommand\bea{\begin{eqnarray}}
\newcommand\eea{\end{eqnarray}}
\newcommand{\nonum}{\nonumber}
\begin{document}

\title{\bf Josephson decoupling phase in 
inhomoheneous arrays of superconducting     
quantum dots}

\author{\bf Sujit Sarkar}
\address{\it PoornaPrajna Institute of Scientific Research,
4 Sadashivanagar, Bangalore 5600 80, India.\\    
}
\date{\today}

\begin{abstract}
We present a study of quantum phase analysis of inhomogeneous and
homogeneous arrays of superconducting quantum dots (SQD). We observe
the existance of Josephson decouple (JD) phase
only at the half filling for inhomogeneous array of
SQD due to the fluctuation of
Josephson couplings over the sites at half filling. 
In JD phase superconductivity
disappears even in the absence of Coulomb blockade phase. 
We also observe that fluctuation of on-site Coulomb charging energy
produces the relevant coupling term that yields
Coulomb blockade gapped phase. 
The presence of nearest-neighbor and next-nearest-neighbor Coulomb
interaction yields the same physics for inhomogeneous and homogeneous SQD. \\  
\end{abstract}

\maketitle


\noindent
{\bf Introduction:} It is well known that the quantal
phase ($ \phi $) of superconductor is coherent over the superconducting 
system; therefore, we expect the quantum properties of the electron 
to be 
visible at a macroscopic level \cite{jose1,jose2,bard1,bard2,coo,pde}. 
The Josephson effect is nothing but the manifestation of coherence of
the superconducting quantal phase in the system. 
In this effect, the system is gaining 
energy to stabilize the superconducting phase. 
A superconducting phase is stable when  
the Josephson coupling ($E_J $) between two 
superconductors separated by a junction, is larger
than the Cooper pair charging energy ($ E_c $). This
is the conventional wisdom in the literature of 
superconductivity \cite{jose1,jose2,bard1,bard2,coo,pde}.
In this letter we raise the question for a nanostructure superconducting
system whether this conventional wisdom is still valid for an 
inhomogeneous SQD system
with fluctuating Josephson coupling.  
We will see that for inhomogeneous SQD array, our model system is in the
insulating phase even in the absence of Coulomb blockade phase. Although
the system has finite $E_J$, we
characterize this phase as a Josephson decouple phase (JD) because it is
not yielding any superconductivity in the system. 
In the present stage our model is completely the theoretical model of
inhomogeneous SQD array which predicts the JD phase. We hope that
the state of engineering of nanoscale superconducting system will
find this type of system. 
Our prime motivation is to predict the JD phase, after the fourty seven
years of the discovery Josephson effect.
We don't think that our model system is
a perfect model of granular superconducting system because we have
built the model to predict the JD phase only 
\cite{and,abe,shapira,jae,orr,sudip,efe,sim}.
In this
study we also raise the question of the effect of on-site Coulomb charging
energy and also fluctuations of it over sites. We will see that this effect
is quite interesting. Apart from that we also study the effect of
nearest-neighbor (NN) and next-nearest-neighbor (NNN) Coulomb interactions
for inhomogeneous and homogeneous SQD.
\\
%
%
{\bf General field-theoretical formalism for inhomogeneous SQD and
homogeneous SQD arrays: }\\
At first we write down the model Hamiltonian of 
inhomogeneous SQD system with 
fluctuating (with a periodicity of two lattice sites, this model is
sufficient to detect the JD phase)
Josephson couplings with on-site charging energies and intersite
interactions in presence of gate voltage. The Hamiltonian is written
as 
\beq
H~=~H_{J1}~+~H_{EC0}~+~H_{EC1}~+~H_{EC2}.
\eeq
We recast the different parts of the Hamiltonian in quantum
phase model as.\\
\begin{center}
$
H_{J1}~=~ -E_{J1} \sum_{i} (1 - {(-1)^i} {\delta}_1) 
cos ({\phi}_{i+1} -{\phi}_{i}),
$
\end{center}
where 
${\phi}_i $ and $ {\phi}_{i+1} $ are quantal phase of the SQD at the point
i and i+1 respectively.
Josephson couplings are fluctuating 
over the sites,
${E_{J1}} (1 + {\delta}_1 )$ and
${E_{J1}} (1 - {\delta}_1 )$ are the Josephson coupling strength
for odd and even site respectively. We also consider the fluctuations 
of on-site Coulomb charging energy over the sites. This is represented
as
\begin{center}
$
H_{EC0}~=~ \frac{E_{C0}}{2} \sum_{i} (1 - (-1)^{i} {\delta}_2 )
{(-i \frac{\partial}{{\partial}{{\phi}_i}} - \frac{N}{2})^{2} }
$,
\end{center}
where ${E_{C0}}$ is the on-site charging energy.
$ E_{C0} ( 1+ {\delta}_2 )$ and
$ E_{C0} ( 1- {\delta}_2 )$ are the on site charging energies for
odd and even sites respectively.
All $\delta$'s are deviations
of exchange couplings from the homogeneous SQD.
$H_{EC1}$ and $H_{EC2}$ are respectively the Hamiltonians for
nearest neighbor(NN) and next-nearest-neighbor(NNN) interations between SQD.
Now
$$
H_{EC1}~=~ E_{Z1} \sum_{i} {n_i}~{n_{i+1}},
$$
and
$$ H_{EC2}~=~ E_{Z2} \sum_{i} {n_i}~{n_{i+2}} ,$$
where
$E_{Z1}$ and $ E_{Z2} $ are respectively the NN and NNN charging 
energies between
the dots. 
We see that this model is sufficient to explain
JD induced gapped phase of the system. 
In the phase representation,  
$(-i \frac{\partial}{{\partial}{{\phi}_i}})$ is the operator
representing the number of Cooper pairs at the ith dot, 
and thus it takes only the integer
values ($n_i$). Hamiltonian $H_{EC0}$ accounts for the
influence of gate voltage ($e N \sim V_g$). 
$e N$ is the average dot charge induced by the gate voltage.
When the
ratio $\frac{E_{J1}}{E_{C0}} \rightarrow 0$, the SQD 
array is in the insulating state having a gap of the width
$\sim {E_{C0}}$, since it costs an energy $\sim E_{C0}$
to change the number of pairs at any dot. The exceptions are the 
discrete points at $N~=~2n+1$, where a dot with charge $2ne$
and $2 (n+1) e$ has the same energy because the gate charge 
compensates the charges of extra Cooper pair in the dot.
On this degeneracy point, a small amount of Josephson coupling 
leads the system to the superconducting state.\\
Here we would like to recast our basic Hamiltonians in the spin 
language. During this process we follow Ref. \cite{ss} and \cite{lar}.
We map this model to the spin chain model when on-site charging
energy is larger Josephson coupling. Now
\begin{center}
$
H_{J1}~=~ -2~E_{J1} \sum_{i} (1 - (-1)^{i} {{\delta}_ 1} )
( {S_i}^{\dagger} {S_{i+1}}^{-} + h.c)
$,
\end{center}
and
\begin{center}
$
H_{EC0}~=~ \frac{E_{C0}}{2} \sum_{i} (1 - (-1)^{i} {{\delta}_ 2} )
{(2 {S_i}^{Z} - h )^{2} }.$
$
H_{EC1}~=~4 E_{Z1} \sum_{i} {S_i}^{Z}~{S_{i+1}}^{Z},
$
$
H_{EC2}~=~4 E_{Z2} \sum_{i} {S_i}^{Z}~{S_{i+2}}^{Z} .
$
\end{center}
Here $h = \frac{N - 2n - 1}{2} $ allows the tuning of the system around the 
degeneracy point by means of gate voltage. Now we use Abelian bosonization
method to solve this problem.
We recast the spinless
fermion operators in terms of field operators by this relation \cite{gia1}: 
\beq
 {\psi}(x)~=~~[e^{i k_F x} ~ {\psi}_{R}(x)~+~e^{-i k_F x} ~ {\psi}_{L}(x)] 
\eeq
where ${\psi}_{R} (x)$ and ${\psi}_{L}(x) $ describe the second-quantized 
fields of right- and 
left-moving fermions respectively and 
$k_F$ is the Fermi wave vector. 
It is revealed from $H_{EC0}$ that the applied external gate voltage on the
dot systems appears as a magnetic field in the spin chain. 
In our system $k_F$ will depend on the
applied gate voltage.
Therefore, one can study the effect of gate voltage 
through arbitrary $k_F$. 
We would like to express the fermionic fields in terms of bosonic 
field by the relation 
$ {{\psi}_{r}} (x)~=~~\frac{U_r}{\sqrt{2 \pi \alpha}}~
~e^{-i ~(r \phi (x)~-~ \theta (x))} $,
$r$ is denoting the chirality of the fermionic fields,
right (1) or left movers (-1).
The operator $U_r$ commutes with the bosonic field. $U_r$ of different species
commute and $U_r$ of the same species anti-commute. 
$\phi$ field corresponds to the 
quantum fluctuations (bosonic) of spin and $\theta$ is the dual field of $\phi$. 
They are
related by the relations 
$ {\phi}_{R}~=~~ \theta ~-~ \phi$ and  $ {\phi}_{L}~=~~ \theta ~+~ \phi$.
After continuum field theoretical studies for arbitrary
values of $k_F$, the model Hamiltonian becomes
\bea
{H_1} &=& {H_0} + 2 \frac{{E_{J1}}}{2 \pi {\alpha}} {{\delta}_1} \int dx
: cos ( 2 \sqrt{K}{\phi} (x) - (2 k_F - \pi)x ):\nonum\\
& &  + \frac{h E_{C0}}{\pi \alpha} \int dx {{\partial}_x} {\phi (x) } \nonum\\
& &  + \frac{2 h E_{C0}{{\delta}_2} }{\pi \alpha} \int (-1)^{x}
: cos( 2 \sqrt{K} {\phi } (x) +2 k_F x ):~ dx \nonum\\
& &  +~\frac{4 E_{Z1}}{{(2 \pi {\alpha})}^2}
 \int : cos( 4 \sqrt{K} {\phi } (x)~- \nonum\\
& & (G - 4 k_F)x -2 k_F a ):~ dx \nonum\\ 
& & + ~\frac{4 E_{Z2}}{{(2 \pi {\alpha})}^2}
 \int : cos( 4 \sqrt{K} {\phi } (x)~+ \nonum\\ 
& & (G - 4 k_F)x - 4 k_F a ):~ dx . \nonum\\
\eea
The Bosonized Hamiltonians for homogeneous SQD can be written as
\bea
{H_2 } &=& {H_0} + \frac{h E_{C0}}{\pi \alpha}  
 \int dx {{\partial}_x} {\phi (x) } \nonum\\ 
& & + \frac{4 E_{Z1}}{{(2 \pi {\alpha})}^2} 
\int : cos( 4 \sqrt{K} {\phi } (x) \nonum\\
& & ~- (G - 4 k_F)x -2 k_F a ):~ dx \nonum\\ 
& & + ~\frac{4 E_{Z2}}{{(2 \pi {\alpha})}^2} \nonum\\
& &  \int : cos( 4 \sqrt{K} {\phi } (x)~+ (G - 4 k_F)x - 4 k_F a ):~ dx , \\ 
\eea
and
\bea
H_0 & = &  ( \frac{v}{2 \pi} + \frac{8 E_{C0}}{{\pi}^2} 
- \frac{2 {E_{J1}}^2}{E_{C0}}) \nonum\\
& &  ~\int dx ~[ 
:{{({{\partial}_x} \theta)}^2}:
+ :{{({{\partial}_x} \phi)}^2}:~] \nonum\\
& & + (16 E_{C0} - 4 \frac{{E_{J1}}^2}{E_{C0}}) \nonum\\ 
& &  \int dx~
: ({\partial}_x {\theta} - {\partial}_x {\phi}) 
({\partial}_x {\theta} + {\partial}_x {\phi}): 
\eea
Here, $H_0 $ is the non-interacting part of the model Hamiltonian, 
$v$ is the velocity of low energy excitations, one of the Luttinger
liquid parameter and the
other is  
$ K $.
And $ G $ is the reciprocal lattice vector.\\

{\it Results  
and physical interpretation: }
\noindent
Here we study the relevant physics for
single Cooper pair in alternate site for inhomogeneous and 
homogeneous SQD system (here ${k_F}= \frac{\pi}{2}$ because the system is 
at half-filling).
The effective Hamiltonian for the inhomogeneous SQD reduce to 
\bea
{H_1} & = & {H_0} + 2 \frac{E_{J1}}{2 \pi \alpha} {{\delta}_1}
\int ~dx :cos(2 \sqrt(K) \phi (x)): \nonum\\
& & + {h E_{C0}} \int  ({{\partial}_x} \phi (x))~dx \nonum\\
& & + 2 \frac{h E_{C0}}{\pi \alpha} {{\delta}_2}
\int ~dx :cos(2 \sqrt(K) \phi (x)): \nonum\\
& & - \frac{4 ( E_{Z1}- E_{Z2})  }{{(2 \pi \alpha)}^2}
\int ~dx :cos(4 \sqrt{K} \phi (x)): . \nonum\\
\eea
Our model Hamiltonian consists of three sine-Gordon couplings.
The second term of the Hamiltonian arises due to fluctuations 
of Josephson
coupling. It yields the gapped phase of the system.
The anamolous scaling dimension of this term is 2$K$.
This phase
is spontaneous, i.e., infinitesimal variation of NN Josephson coupling
around sites
is sufficient to produce this state.
When $ {E_c }$ is larger than $ {E_J}$ 
the system is in the gapped phase due to the Coulomb
blockade effect. If we consider the case when $E_J$ is much smaller than 
$ E_C $ then one should naively think that the system is in the superconducting
phase but the situation here is quite different due to the fluctuations 
of Josephson coupling, its produces the gap state in the system
and blocks the superconducting phase of the system.
We term this phase as Josephson decoupling phase because it is
not yielding any superconducting phase due to the tunneling at
different SQD; this phase is present
even in the absence of Coulomb blockade.
This gapped state 
prevails until the applied gate voltage is sufficient to break
this gapped phase \cite{cbl}.
This prediction is absent in all previous studies of superconductivity 
\cite{jose1,jose2,pde,bard1,bard2,coo,ss,lar,lik1,lik2,baro}. 
The third term of the Hamiltonian arises due to
constant Coulomb charging energy; it promotes the system in
different charge quantized state due to the variation of applied
gate voltage. The fourth term of the Hamiltonian is due to the 
fluctuations of on-site Coulomb charging energy. It is like
the staggered magnetization of the system.
It's anamolous scaling dimension is
also $ 2 K $. Therefore, the system is in the mixed gapped state when both
terms are present. The fourth term arises due to the NN and NNN interactions;
the anamolous scaling dimension of this term is $4 K $. Therefore the physics
of gapped state is mainly governed by the second and the fourth term of the 
Hamiltonian.\\

Effective Hamiltonian for homogeneous SQD array is
\bea
{H_2} & = & H_0 ~-~\frac{4 ( E_{Z1} - E_{Z2} )}{{(2 \pi {\alpha})}^2}
 \int cos( 4 \sqrt{K} {\phi } (x)~)~ dx \nonum\\
& & +
\frac{ h E_{C0}}{\pi \alpha} \int ~{{\partial}_x}{\phi}~dx.
\eea
When $ E_{Z2} $ exceed some critical value,
the ground state of the system is dimerized and doubly degenerates. The dimerized
ground state is the product of spin singlet of adjacent sites \cite{hal}.
When $ E_{Z2} $ is less than a critical value the physics of the
system is governed by the $ E_{Z1} $ and the gapped phase of the system is
alike to spin-fluid phase of the system. In this model Hamiltonian, there is
no relevant sine-Gordon coupling term present due to the variation of 
Josephson coupling. Therefore there is no JD phase for homogeneous SQD. 
We also study
our model Hamiltonian for different densities (by varying $k_F $) 
but we are unable to find
JB phase for any other fillings for both inhomogeneous and homogeneous
SQD array.\\
{\bf Conclusions}:
We have predicted the evidence for the Josephson decouple phase for
inhomogeneous SQD only at half-fillings. This is the first 
prediction of Josephson decoupled phase  
in the literature for these type of system. There is no
evidence of Josephson decouple phase for homogeneous SQD. We
have also predicted the interesting behavior of the system
due to the fluctuating on-site Coulomb charging energy. Our
prediction of Josephson decoupling phase after the fourty
seven years of Josephson effect; we hope that evidence of this
JD phase will be verified experimentally as the Josephson effect
has verified experimentally after the theoretical prediction. 


\begin{references}
\bibitem{cbl} Here we discuss the effect of gate voltage on 
homogeneous and non-homogeneous SQD 
and the charge quantization physics.
From an energetic point of view, when an Cooper pair added to SQD,
the new electrostatic energy with added Cooper pair is smaller than
the corresponding energy in the absence of Cooper pair
\beq
\frac{1}{2 C} {( |Q|- |2 e|)}^{2} \leq ~\frac{Q^2}{2 C} => Q \geq |2e|.
\eeq
Therefore we need a bias $ |V| \geq \frac{2e}{2C} $ for current to flow.
Hence Cooper pair transport is unfavorable when $ |V| \leq \frac{2e}{2C} $,
this is the Coulomb blocked phase for non-homogeneous and homogeneous SQD. 
When we added an extra Cooper pair
to the SQD, we expect a spikes in conductance, i.e, the steps in correspondence
to the biases that overcome charging energies with increasing number of
Cooper pair. When one kept fixed the bias and gate voltage ($ V_g $ ) applied
to SQD (with capacitance $ C_g $ ). The 
applied gate voltage will vary densities of Cooper pair in SQD. In presence
of N Cooper pair in SQD, the electrostatic energy
$ {E_N} = \frac{ {(2 N e)}^{2} }{2 e_g } - 2 {V_g } N |e| $. Adding of extra
Cooper pair is favorable when the above energy is minimized w.r.t number of
Cooper pair, it implies $ |V_g | = \frac{|2 e|}{C_g } (N + 1/2) $. At these
gate voltage the current has a discontinuity. Therefore we define the 
charge quantized state when the applied gate voltage
is away from the optimum gate voltage point and also for 
$ | V| < \frac{2 |e|}{2 c} $.
\end{references}
\end{document}